\documentclass[3p]{elsarticle}

\usepackage{tikz}
\usepackage[english]{babel}
\usepackage{siunitx}
\usepackage[fleqn]{amsmath}
\usepackage{my_abrev}
\usepackage[
  colorlinks=true,
  citecolor=blue,
  urlcolor=blue,
  linkcolor=blue
]{hyperref}
\usepackage[autostyle=true,german=quotes]{csquotes}
\usepackage{pgfplots}
\usepackage{subcaption}
\usetikzlibrary{external}
\biboptions{semicolon,authoryear}

\begin{document}
\begin{frontmatter}
\title{Reduced order modelling for spatial-temporal temperature and property estimation in a multi-stage hot sheet metal forming process}

\author[1]{Daniel Kloeser}
\author[2]{Juri Martschin}
\author[1]{Thomas Meurer}
\author[2]{Erman Tekkaya}

\address[1]{Chair of Automatic Control,	Faculty of Engineering, Kiel University, Kiel, Germany.}
\address[2]{Institute of Forming Technology and Lightweight
  Components, Technische Universität Dortmund, Dortmund,Germany }

\begin{abstract}
  A concise approach is proposed to determine a reduced order
  control design oriented dynamical model of a multi-stage hot sheet
  metal forming process starting from a high-dimensional coupled
  thermo-mechanical model. The obtained reduced order nonlinear
  parametric model serves as basis for the design of an Extended
  Kalman filter to estimate the spatial-temporal temperature
  distribution in the sheet metal blank during the forming process
  based on sparse local temperature measurements. To address modeling
  and approximation errors and to capture physical effects neglected
  during the approximation such as phase transformation from austenite to martensite a disturbance
  model is integrated into the Kalman filter to achieve joint state
  and disturbance estimation. The extension to spatial-temporal
  property estimation is introduced. The approach is evaluated for a
  hole-flanging process using a thermo-mechanical simulation model evaluated
  using \lsdyna. Here, the number of states is reduced from
  approximately \num{17000} to \num{30} while preserving the relevant
  dynamics and the computational time is 1000 times shorter. The performance of the combined temperature and
  disturbance estimation is validated in different simulation scenarios
  with three spatially fixed temperature measurements.
\end{abstract}


\end{frontmatter}

\section{Introduction}
\label{sec:introduction}

Multi-stage hot sheet metal forming enables the production of complex
formed press-hardened components at high stroke rates
\citep{Belanger2017}. The sheet metal blank is first heated in a
furnace, then optionally pre-cooled to the desired initial forming
temperature and finally formed and quenched in a sequence of tools in
a transfer die \citep{Belanger2016}. Several different forming
operations are carried out on a component before the temperature falls
below a critical temperature, e.g., the martensite start
temperature. \citet{Lobbe2016} presented a process design for a
progressive die in which the sheet metal coil is pre-punched, then
rapidly austenitized by means of induction heating and formed and
quenched in several consecutive stages. A modification of this
process chain is the implementation of hot stamping operations in
progressive die plate forging of tailored high strength gear parts by
partial resistance heating \citep{Mori2017}.  

These production methods have been limited to special applications due
to the partially unknown and difficult to predict 
interplay of thermal and mechanical influences throughout the process
chain. For example, the deformation history during hot forming is
affected by the microstructure development and also influences the
microstructure development itself \citep{Nikravesh2012}. A forming and
heat treatment depending shift of the transformation point from
austenite to martensite causes a change in the process chain
temperature history, which affects the forming process through a
change in the temperature- and microstructure-dependent flow stress
\citep{Venturato2017}. Besides the thermal-mechanical interactions,
there are difficult to assess service life depended process influences
due to batch fluctuations and the varying tool condition
\citep{Gracia-Escosa2017}. The given complexity and the lack of
suitable prediction models complicate setting and maintaining
desired product properties such as geometry and hardness.

In metal forming processes, models are usually obtained using the
finite element method (FEM) within the design step. The 
resulting models are accurate but not real-time capable due to high
dimensions and nonlinearities of the system. Hence a major challenge for
estimator design is to develop models of reduced complexity and
dimension that are still sufficiently accurate.
This can be achieved by model order reduction techniques that enable
us to reduce model dimension while preserving the relevant system
dynamics  (see, e.g., the treatise in \cite{antoulas:05}). 
Parametric model order reduction techniques provide a suitable
approximation of the full-order dynamical system over a range of
parameters (see, e.g., the review in \cite{Benner2015}). This approach has been
successfully applied to a broad range of applications including
structural dynamics, aeroelastic models as well as electrochemical and
electro-thermal applications. While different approaches can be
applied to compute the \glspl{rom} subsequently
projection-based model order reduction is considered using \gls{pod},
which can be applied to linear as well as nonlinear
models. Basis vectors used for projection in \gls{pod} are typically
computed using the method of snapshots
\citep{Sirovich1987}. \cite{Benner2015} provide a rather comprehensive
survey on its development, variants and selected applications. In the context of
sheet metal forming the application of model order 
reduction techniques and in particular \gls{pod} can be found, e.g.,
in \cite{bolzon:EffectiveComputationalTool:2011,radermacher:ModelReductionElastoplasticity:2014} for
a (quasi-static) elastoplastic mechanical problems. \citet{Bohm2017}
compare several linear model reduction techniques for the temperature 
control of deep drawing tools and develop a trajectory planning and
feedforward control strategy based on the obtained \gls{rom}. 

To enable robust and versatile production, a closed loop control has
to be employed \citep{Allwood2016a}, where product properties as well as
decisive process variables are measured in situ during the multi-stage
forming and heat treatment process. The measurements are fed back to a
controller and set by adjusting the process parameters using real-time
capable models. Possible actuation parameters include the kinematics
of the tool and the austenitization parameters. 
The influence of process parameters on product quality in a press
hardening application is considered in
\cite{landgrebe_etal:2015}. Related results are provided in
\cite{wang_etal:2017} by taking into account the control of dwell
pressure or dwell time, respectively. In both cases the preliminary
heating is not considered as control variable. 
Models for controlling springback \citep{Lobbe2015}, microstructure \citep{Lobbe2016} and strength
\citep{Lobbe2018a} in heat-assisted bending of
sheet material in a progressive die are available. However, it is not yet possible to implement
multivariable control of geometric and mechanical or 
microstructural properties in multi-stage hot sheet metal
forming. Knowledge of the temporally and spatially varying temperature
distribution of the sheet is substantial for controlling the given
process chain. In particular the temperature evolution is decisive
for the development of the microstructure and the deformation
behaviour of the material.
In order to reconstruct the spatial-temporal temperature distribution
from sparsely available measurement data such as pyrometers,
model-based state estimation methods are inevitable. \citet{Speicher2014} designed an
\gls{ekf} for the plate rolling process and \citet{Zheng2011} for the
strip rolling, respectively. Force measurements are used in
\cite{havinga:PRODUCTTOPRODUCTSTATEESTIMATION:2017} to estimate
product properties by exploiting an interpolation process model
obtained from a finite element model. Herein the structured
uncertainties are addressed using a bias model. Taking into account
tools from robust nonlinear control the design of a robust observer in view of
additive parametric uncertainties is studied in
\cite{benosman:DatadrivenRobustState:2021a} based on \glspl{rom} for
the discretized 2D Boussinesq equations to estimate air flow and
temperature distributions for heating, ventilation, and air
conditioning management.  

In this work, a concise approach is presented that enables us to
systematically determine a \gls{rom} suitable for estimator
design based on the thermo-mechanical high dimensional finite element model for a
multi-stage hot forming process. The proposed approach is based on the
assumption of a one-sided decoupling between the thermal and the 
mechanical subsystems. This results in a parametric nonlinear thermal
model that is utilized for model order reduction using
\gls{pod}. The obtained \gls{rom} is exploited for the design of an
\gls{ekf} to obtain a real-time estimation of the spatial-temporal 
temperature distribution of the sheet metal blank during the forming
process based on only a few selected local temperature
measurements. This furthermore enables us to deduce
temperature-dependent spatial-temporal material properties, e.g.,
hardness, by taking into account suitable constitutive
relationships. To simplify numerical evaluation of the \gls{ekf} a
model simplification involving a so-called supporting or reference 
trajectory is proposed to obtain a linear time-varying model as a special form of
a parametric model. To account for modeling errors, e.g., thermal
effects of the phase transition, and reduction errors the setup is
amended by a disturbance estimation that enables us to improve the
estimator performance. The developments are validated in 
numerical simulations for a hole-flanging process.  

The paper is structured as
follows. Section~\ref{sec:problem_formulation} provides the problem
formulation and the classical design-oriented simulation model for the
coupled thermo-plastic forming process. In
Section~\ref{sec:reduced_order_modelling} a parametric \gls{rom} is
deduced, which is the starting point for the \gls{ekf} design in
Section~\ref{sec:state_and_disturbance_estimation}. Herein, trajectory-oriented linearization to obtain a linear time-varying model is
considered and a disturbance estimation is integrated into the \gls{ekf}. The 
methods are evaluated in Section~\ref{sec:simulation_results} for
hole-flanging process. Section~\ref{sec:conclusion} concludes this
paper.

\section{Problem formulation}
\label{sec:problem_formulation}

In the following, the multi-stage hot forming process is introduced
and the necessity for an estimator to reconstruct online the spatial-temporal temperature
distribution is motivated. The 
section concludes with a description of the fully coupled \lsdyna
simulation which is used as starting point for the model order
reduction.  

\subsection{Multi-stage forming process}

\begin{figure}[!ht]
  \centering
  \includegraphics{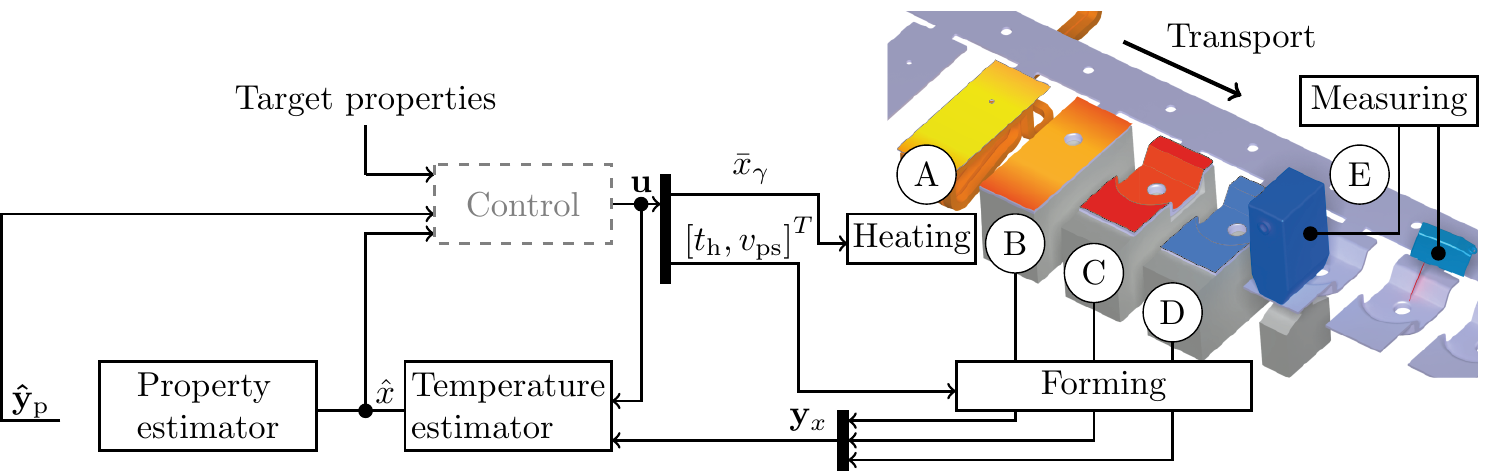}
  \caption{Schematics of estimator-based closed-loop property
    control in the progressive die.}
  \label{fig:process}
\end{figure}

The multi-stage hot sheet metal forming demonstrator process
illustrated in Figure~\ref{fig:process} for a progressive die is
considered subsequently as introduced in \citet{Lobbe2015}.
After pre-punching the coil, the sheet metal blank is homogeneously
heated to the austenitizing temperature \tempaust by a combination of
inductive and conductive heating (A). In stages (B) to (D) a sequence
of hole-flanging (B), combined deep drawing and stretch drawing (C)
as well as die bending (D) is carried out. At the same time, quenching
takes place through tool contact.
Figure \ref{fig:process} in addition shows the process inputs and the
integration of the subsequently determined temperature estimator to be used
for both process monitoring and model-based property oriented feedback
control.
The process input variables that may be eventually used to control the
forming process read
\begin{align}
  \label{eq:process:inputs}
  \controlvec=
  \begin{bmatrix}
    \tempaustavg & \punchspeed & \holdingtime
  \end{bmatrix}^T.
\end{align}
Herein, $\tempaustavg$ is the average austenitizing temperature, which
is adjusted by an underlying control loop, $\punchspeed$ is the punch
speed, and $\holdingtime$ is the holding time at the bottom dead
point. Obviously, $\punchspeed$ and $\holdingtime$ refer to the
kinematics of the ram and hence influence directly the forming steps.  

In stages (A) to (D) of the real process temperature measurements 
$\meastemperature(\timevar)$ are recorded at certain time instances by thermal
imaging cameras or pyrometers. Before the individual components are
separated from the strip boundary, the geometry and selected product
properties $\measproductproperties(\timevar)$ such as the microstructure are
determined using a laser scanner and a 3MA-system, respectively, in
stage (E).  

For the design of the temperature estimation the local temperature measurements
$\meastemperature(\timevar)$ from stages (A) to (D) serve as output injections
into the estimator equations to reconstruct the spatial-temporal
temperature distribution $\hat{\temperature}(\nodeloc,\timevar)$ on
the sheet metal blank. Based on this estimate we aim to assess and to
evaluate the spatial-temporal distribution of key product properties $\mathbf{\hat{y}}_{\text{p}}(\timevar)$
such as hardness, yield stress or geometry that dependent on the
temperature and deformation history with the property estimator. This paper focuses on the design
and evaluation of the spatial-temporal robust temperature
distribution and outlines the property estimation. The latter is to be
combined with the property characterization considered in
\cite{martschin_etal_metals:21} towards the development of model-based
property control in subsequent work.  

\subsection{Fully coupled simulation}

For the numerical simulation of the sheet metal blank temperature
distribution during the multi-stage process, the forming steps, the
heat-exchange between sheet, tool as well as the environment and the
thermo-mechanical interplay are modeled in accordance with
\citet{Hochholdinger2011}. Hereby the FE code \lsdyna (solver:
R11.1.0) is used. Tools are modeled with rigid and the blank with
elasto-viscoplastic 12 node thermal thick shells. Characteristic
values for convection and radiation are chosen based on
\citet{Shapiro2009}. Martensitic stainless steel \xchrome%
is used in the
simulation. Compared to conventional press hardening steels, e.g.,
22MnB5, the low martensite start temperature of approximately
\SI{443}{\kelvin} \citep{Dieck2017} allows a longer period of time for
the multi-stage forming. %
The material behaviour of \xchrome during hot
forming is modeled by a simplified adaptation of the material model
Mat 248 presented by \citet{Hippchen2016}. Temperature and strain
rate-dependent flow curves and Young’s modulus, thermal conductivity,
heat capacity, density, and thermal expansion parameters are based on
\citet{Spittel2009}. Only a phase transformation from austenite to
martensite is considered, since higher cooling rates are achieved in
the progressive die than in air (air-hardening steel). The
austenitizing of the sheet by combined inductive and conductive
heating is not simulated. Prior to the transfer into the first forming
stage, a fully austenitized sheet with a defined temperature
distribution is assumed.

\section{Reduced order modelling}
\label{sec:reduced_order_modelling}

The complexity of the \lsdyna model is generally high and not suitable
for online estimation purposes. Therefore, it is simplified in the
following steps. At first, the thermal solution is separated from the
mechanical solution assuming a one-sided decoupling. Secondly,
ordinary differential equations are derived from the partial
differential equation governing heat conduction using the finite
element method so that a parameter- and state-dependent system is
obtained. Third, a \gls{rom} is determined using \gls{pod}.

\subsection{One-sided decoupling of mechanical and thermal solution}
\label{subsec:one-sided_decoupling_of_mechanical_and_thermal_solution}

To determine an explicit dynamical model of the process dynamics the
fundamental assumption is imposed that the thermal solution can be
separated from the mechanical solution.
\begin{assumption}
  \label{assump:separation}
  Thermal and mechanical subsystem are coupled partially with the
  mechanical solution entering the thermal subsystem but not vice
  versa. 
\end{assumption}
This assumption simplifies the setting but introduces a modeling error
that is evaluated in a simulation study in Section
\ref{sec:simulation_results} for a hole-flanging process.
Hence, only the spatial-temporal 
temperature distribution is determined (solved) online while the
mechanical solution, that impacts the thermal simulation, is computed
offline. The latter enters the thermal subsystem in terms of the
parameter vector $\parametervec(\nodeloc,\timevar,\controlvec)$, which
depends on the process conditions and hence the process inputs
\eqref{eq:process:inputs} and is composed of the following components:
\begin{itemize}
\item $\deformation(\nodeloc,\timevar,\controlvec)$, i.e., the
  deformation of the sheet metal blank with respect to the initial
  shape at location $\nodeloc$; 
\item $\tooldist(\nodeloc,\timevar,\controlvec)$, i.e. the
  shortest distance between the sheet metal blank and the tools at
  location $\nodeloc$; 
\item $\pressure(\nodeloc,\timevar,\controlvec)$, i.e., the
  contact pressure between the sheet metal blank and the tool at
  location $\nodeloc$; 
\item $\tempinf(\nodeloc,\timevar,\controlvec)$, i.e., the
  contact temperature at location $\nodeloc$. This is either the
  ambient temperature if $\tooldist(\nodeloc,\timevar)$ is larger than
  a certain threshold or the tool temperature if it is smaller. 
\end{itemize}
If clear from the context we subsequently write $\parametervec$ and
avoid the individual dependencies of the parameters. 

\subsection{Continuum formulation}
\label{subsec:continuum_formulation}

Modelling of the temperature distribution
$\temperature(\nodeloc,\timevar)$ on the deforming sheet metal blank,
whose shape at time $\timevar>0$ is determined by the volume
$\volume(\timevar)\subset\mathbb{R}^3$ with boundary surface $\surface(\timevar)$ starting from the undeformed
shape $\volume(0)=\volume_0$, $\surface(0)=\surface_0$, leads to the
heat equation with in general temperature dependent
material properties
\begin{subequations}
  \label{eq:heat_eq}
  \begin{align}
    \label{eq:heat_eq_pde}
    \density \heatcap[(\temperature)]\partial_\timevar
    \temperature&=\divop\left[\heatcond[(\temperature)]\gradop
                                     \temperature\right]+\heatinduced(\temperature),
    &(\nodeloc,\timevar)\in\volume(\timevar)\times\mathbb{R}^{+}_{0},\\
    \label{eq:heat_eq_init}
    \temperature\vert_{\timevar=0}&=\tempaust, &\nodeloc \in\volume_0\cup\surface_0, \\
    \label{eq:heat_eq_bound}
    \heatcond(\temperature)\gradop \temperature\cdot\surfacenormal&=\heatcondbound(\temperature,\parametervec)\left[\tempinf-\temperature\right], &(\nodeloc,\timevar) \in \surface(\timevar)\times\mathbb{R}^{+}_{0}.
  \end{align}
\end{subequations}
The material density \density is assumed constant while the specific heat
$\heatcap(\temperature(\nodeloc,\timevar))$ and the thermal
conductivity $\heatcond(\temperature(\nodeloc,\timevar))$ depend on
the temperature distribution. The parameter
$\heatinduced(\temperature(\nodeloc,\timevar))$ refers to the induced
heat and can represent, e.g., thermal effects of the phase
transformation.  
The initial condition \eqref{eq:heat_eq_init} is given by the
austenitizing temperature $\tempaust(\nodeloc)$ which is set by the
inductive heating process. In the following we assume that
$\tempaust(\nodeloc)=\tempaustavg$, i.e., the austenitizing
temperature is homogeneously distributed.
The boundary conditions~\eqref{eq:heat_eq_bound} in the direction of the outer normal $\surfacenormal(\timevar)$ of the boundary surface $\surface(\timevar)$ are modelled by mixed boundary conditions. The values for the thermal transfer coefficient $\heatcondbound(\temperature(\nodeloc,\timevar), \parametervec)$ and the contact temperature $\tempinf(\nodeloc,\timevar,\controlvec)$ are determined in accordance to \lsdyna \citep{LivermoreSoftwareTechnologyCorporation2007,Hochholdinger2011}.
The parameter vector $\parametervec$ with the mechanical solution
enters the formulation in two ways: (i) the time evolution of
$\volume(\timevar)$ depends on $\volume_0$, i.e., the undeformed
state, and the deformation $\deformation(\nodeloc,\timevar,\controlvec)$, (ii) the other
parameters in \parametervec influence \eqref{eq:heat_eq_bound}. 
\begin{remark}[Material derivative]
  Let $D_t$ refer to the material derivative. The system
  representation \eqref{eq:heat_eq} is based on the assumption that 
  \begin{align*}
    D_t\temperature(\nodeloc,\timevar) = \partial_\timevar
  \temperature(\nodeloc,\timevar) + \nabla \cdot (\vec{v}(\nodeloc,\timevar)
    \temperature(\nodeloc,\timevar))
    \approx \partial_\timevar\temperature(\nodeloc,\timevar)
  \end{align*}
  with $\vec{v}(\nodeloc,\timevar)$ the velocity of the domain point
  at $(\nodeloc,\timevar)$. In other words we consider the rate of
  change observed when moving with the particle to be approximately
  identical with the rate of change at a fixed point. 
  However, the setup can be easily generalized by appropriately
  replacing the partial differentiation
  $\partial_\timevar\temperature(\nodeloc,\timevar)$ in
  \eqref{eq:heat_eq_pde} by $D_t\temperature(\nodeloc,\timevar)$
  and taking into account the continuity equation.
\end{remark}
\begin{remark}[Lagrangian viewpoint]
 It seems reasonable to reconsider \eqref{eq:heat_eq} from a
 Lagrangian viewpoint as the time evolution of any material point due
 to deformation is known a-priori by means of the separation
 introduced in Assumption \ref{assump:separation}. This, however,
 introduces issues when considering the temporary contact between tool
 and sheet  during forming. Moreover, this requires to approximate the
 Jacobian matrix between material and spatial coordinates online and
 in principle to add the differential equations determining the
 temporal dynamics of the material points. A comparison between the
 considered approach and a Lagrangian approach is not at the core of
 this paper and is omitted. 
\end{remark}

\subsection{Weak formulation}

The distributed parameter description~\eqref{eq:heat_eq} can be recast
into a weak formulation using a test function $\weightfun(\cdot,\timevar)\in
H^1(\volume(\timevar))$ for any fixed but arbitrary $\timevar\in\mathbb{R}^+_0$.
Then multiplication of \eqref{eq:heat_eq_pde} with the test function
and integration over the domain $\volume(\timevar)$ provides
\begin{align}
  \density \int_{\volume(\timevar)}
  \weightfun\heatcap(\temperature) \partial_\timevar
  \temperature \intd\volume= \int_{\volume(\timevar)}
  \weightfun\divop\left[\heatcond[(\temperature)]\gradop
                                     \temperature\right] \intd\volume +
  \int_{\volume(\timevar)} \weightfun
  \heatinduced(\temperature)\intd\volume.
\end{align}
Utilizing integration by parts and incorporating the mixed boundary conditions lead to the weak formulation
\begin{align}
  \label{eq:heat_equation_weak}
  \density \int_{\volume(\timevar)} \weightfun
  \heatcap(\temperature) \partial_\timevar
  \temperature \intd\volume
  & =
    \int_{\surface(\timevar)}
    \weightfun\heatcond(\temperature)\nabla\temperature\cdot\surfacenormal
    \intd\surface
    -
    \int_{\volume(\timevar)}
    \nabla\weightfun\cdot(\heatcond(\temperature)\nabla\temperature) \intd
    \volume\nonumber\\
  &\phantom{=}
    +
    \int_{\volume(\timevar)}
    \weightfun
    \heatinduced(\temperature)
    \intd\volume\nonumber\\
  &=
    \int_{\surface(\timevar)}
    \weightfun\heatcondbound(\temperature,\parametervec)\left[\tempinf-\temperature\right]\intd\surface
    -
    \int_{\volume(\timevar)}
    \nabla\weightfun\cdot(\heatcond(\temperature)\nabla\temperature) \intd
    \volume\nonumber\\
  &\phantom{=}
    +
    \int_{\volume(\timevar)}
    \weightfun
    \heatinduced(\temperature)
    \intd\volume.
\end{align}

\subsection{Full order system formulation using finite element approximation}
\label{subsec:finite_element_method}

The weak formulation in~\eqref{eq:heat_equation_weak} is discretized
in space with the finite element method in order to obtain ordinary
differential equations \citep{Zienkiewicz2013}. For this purpose the
software tool Firedrake
\citep{Rathgeber2016,Balay1997,petsc-user-ref,Dalcin2011,McRae2016,MUMPS01,MUMPS02}
is used taking into account the weak formulation
\eqref{eq:heat_equation_weak}. The resulting system can be written in matrix form 
\begin{align}
  \label{eq:heat_equation_matrix}
  \masmat(\tempvec,\parametervec,\timevar)\tempvecdot
  &=
    \stfmat(\tempvec,\parametervec,\timevar)
    \tempvec+\cnstvec(\tempvec,\parametervec,\timevar), && t>0,\quad \tempvec(0)=\tempvec_{0}\in\realnumber^{\dimstate}
\end{align}
with the in general sparse mass and stiffness matrices
$\masmat(\tempvec,\parametervec,\timevar),~\stfmat(\tempvec,\parametervec,\timevar)\in 
  \realnumber^{\dimstate\times\dimstate}$, respectively. The
inhomogeneity ${\cnstvec (\tempvec,\parametervec,\timevar)\in \realnumber^{\dimstate}}$
results from the mixed boundary conditions~\eqref{eq:heat_eq_bound}
and the induced heat
$\heatinduced(\temperature(\nodeloc,\timevar))$. The explicit
time dependency of the system matrices and the inhomogeneity originates from the
time-varying domain $\volume(\timevar)$ with boundary
$\surface(\timevar)$. The state vector
$\tempvec(\timevar)\in\realnumber^{\dimstate}$ denotes the temperature
at each node of the mesh.
Local temperature measurements are summarized in the output vector
$\measvec_\temperature(\timevar)\in\realnumber^{\dimmeas}$ governed by
\begin{align}
  \label{eq:heat_output}
  \measvec_\temperature
  &=
    \measmat(\timevar)\tempvec,&&t\geq 0.
\end{align}
The output matrix $\measmat(\timevar)\in\realnumber^{\dimmeas\times\dimmeas}$ is time-variant as the
measured temperatures on the sheet metal blank and thus the mapping to
the output vector change during the forming process as the sensors are
usually attached to the tools so that different locations on the blank
are observed during deformation. Moreover, sensors might be shadowed
by the tools at some stages of the forming process.  

\subsection{Model order reduction}
\label{subsec:model_order_reduction_using_pod}

System \eqref{eq:heat_equation_matrix} is composed of several
thousand states and is hence not suitable for estimation and control
purposes. Therefore, model order reduction is considered to systematically reduce
the number of states while preserving the most important dynamics of
the system in the \gls{rom}. The goal of many model order reduction techniques is
to project the states of the original system $\tempvec(\timevar)$
onto a reduced order state space by means of 
\begin{align}
  \label{eq:pod_red_state}
  \tempvec\approx\projectionleft\tempvecred,
\end{align}
where $\tempvecred\cts \in \realnumber^{\dimstatered}$ contains the
reduced states and $\projectionleft \in
\realnumber^{\dimstate\times\dimstatered}$ is the orthonormal
projection matrix.  

A way to choose the basis vector of the projection matrix
$\projectionleft$ is \gls{pod}. The \gls{pod} basis vectors or
\gls{pod} modes, respectively, are chosen
empirically using the method of snapshots \citep{Sirovich1987}. Let
$\tempvec(\timevar;\controlvec)$ denote the solution to
\eqref{eq:heat_equation_matrix} at time $\timevar$ for given initial
state $\tempvec_0$ and process input $\controlvec$ defined in
\eqref{eq:process:inputs}. Note that $\controlvec$ enters the system
description by means of the parameter vector
$\parametervec(\nodeloc,\timevar,\controlvec)$ and covers also the
average austenitizing temperature $\tempaustavg$ that defines the
initial state $\tempvec_0$ according to \eqref{eq:heat_eq_init}. 
With this the snapshot matrix
\begin{align}
  \label{eq:snapshotmatrix}
  \snapshotmatrix=\left[\tempvec(\timevar_1;\controlvec_1),\
  \tempvec(\timevar_2;\controlvec_{2}) ,\ldots,\
  \tempvec(\timevar_{\dimsamples};\controlvec_{\dimsamples})\right] 
\end{align}
with $\snapshotmatrix \in \realnumber^{\dimstate\times\dimsamples}$ is
defined that contains the \dimsamples state solutions
$\tempvec(\timevar_j;\controlvec_j)$
of~\eqref{eq:heat_equation_matrix} for the input $\controlvec_i$ at
time $\timevar_i$. 
The \gls{pod} modes are constructed by using a singular value
decomposition of the snapshot matrix
\begin{align}
  \snapshotmatrix=\svdleft\singvalmat\svdright^T,
\end{align}
where $\svdleft\in\realnumber^{\dimstate\times\dimsamples}$ and
$\svdright\in\realnumber^{\dimstate\times\dimsamples}$ are matrices
composed of the left and right singular vectors of $\snapshotmatrix$
and $\singvalmat\in\realnumber^{\dimsamples\times\dimsamples}$ is the 
diagonal matrix containing the singular values
$\sigma_1\geq\sigma_2\geq...\geq\sigma_{\dimsamples}>0$. The
projection matrix $\projectionleft$ is composed of the $\dimstatered$
column vectors $\svdleft_j$ of $\svdleft$ corresponding to the $\dimstatered$
largest singular values, i.e.,
\begin{align}
  \label{eq:matrix:phi}
  \projectionleft =
  \begin{bmatrix}
    \svdleft_1 & \svdleft_2 & \cdots & \svdleft_{\dimstatered} 
  \end{bmatrix}.
\end{align}
Since the snapshots matrix $\snapshotmatrix$ is
large, a truncated singular value decomposition is performed by making
use of the python library scipy \citep{Virtanen2020}. The singular
values can also give guidance to quantify the number of basis
vectors that are required to obtain a suitably accurate reconstruction
of the snapshots. For this, the threshold
\begin{align}
  \label{eq:pod_energy}
  \podenergy=\frac{\sum_{i=1}^{\dimstatered}\singval_i}{\sum_{i=1}^{\dimsamples}\singval_i}<\podthresh
\end{align}
is considered, where $\podthresh$ is a tolerance specified by the user
and \podenergy is often referred to as the energy of the snapshots
captured by the POD modes.
Utilizing the Galerkin-projection \citep{Benner2015}, the reduced
system reads
\begin{subequations}
  \label{eq:pod_galerkin}
  \begin{align}
    \label{eq:pod_red_system}
    \masmat_\text{r}(\tempvecred,\parametervec,\timevar)\dot{\tempvecred}
    &=
      \stfmat_\text{r}(\tempvecred,\parametervec,\timevar)\tempvecred+\cnstvec_\text{r}(\tempvecred,\parametervec,\timevar),
      && t>0,\quad \tempvecred(0)=\tempvecred_0\in\realnumber^{\dimstatered}
    \\ 
    \measvec_\temperature&=\measmat_\text{r}(\timevar)\tempvecred,&&t\geq 0
  \end{align}
\end{subequations}
with the transformed matrices
\begin{align*}
  &\masmat_\text{r}(\tempvecred,\parametervec,\timevar)
  = \projectionleft^T
  \masmat(\projectionleft\tempvecred,\parametervec,\timevar)
  \projectionleft,&&
  \stfmat_\text{r}(\tempvecred,\parametervec,\timevar)
  =\projectionleft^T \stfmat(\projectionleft\tempvecred,\parametervec,\timevar)
  \projectionleft,\\
  &\cnstvec_\text{r}(\tempvecred,\parametervec,\timevar)
  =\projectionleft^T
  \cnstvec(\projectionleft\tempvecred,\parametervec,\timevar),&&
  \measmat_\text{r}(\timevar)=\measmat(\timevar)\projectionleft.
\end{align*}
The states $\tempvec\cts$ of the full system can be recovered from $\tempvecred\cts$ using \eqref{eq:pod_red_state}.
Note that the POD method is optimal in the sense that it minimizes the
least square error of the snapshot reconstruction but not that it
optimally reconstructs the full model \citep{Rathinam2003}. It is
therefore crucial to select snapshots that excite any relevant
dynamics of the system. 

\section{State, property and disturbance estimation}
\label{sec:state_and_disturbance_estimation}

During the process it is not possible to measure the entire
temperature distribution $\temperature(\nodeloc,\timevar)$ of the
sheet metal blank at any time. Therefore, a state estimation concept
is designed that 
reconstructs online the spatial-temporal temperature distribution
based on the measurements $\measvec_\temperature\cts$, the
parameter vector $\parametervec(\nodeloc,\timevar,\controlvec)$
depending on the solution of the mechanical subsystem and the process
inputs $\controlvec\cts$ as well as the \gls{rom}
\eqref{eq:pod_galerkin}. In addition, two modifications are introduced
that reduce the computational complexity and enable us to estimate the
thermal effect of the phase transformation, which is in general
difficult to include into the model and to parametrize appropriately.
\begin{remark}
  In the following it is assumed that observability or detectability
  of the \gls{rom} is fulfilled. The proper verification of these
  system properties requires to study the obtained models in further
  detail and hence relies on the considered process and the available
  sensors and their placement. 
\end{remark}

\subsection{Extended Kalman filter for temperature and property estimation}
\label{subsec:ekf}

A widely used state and parameter estimation approach is the Kalman
filter \citep{Anderson2012}. In the linear case it provides the
optimal state estimation by the minimizing the covariance 
of the estimation error, i.e., the difference between the estimated state and the
system (model) state under the assumptions of offset-free white
process and measurement noise $\statenoise(\timevar)$ and
$\measnoise(\timevar)$ with the positive definite covariance matrices
$\statenoisemat$ and $\measnoisemat$, respectively. Taking these terms
into account \eqref{eq:pod_galerkin} reads
\begin{subequations}
  \label{eq:ekf_system}
  \begin{align}
    \masmat_\text{r}(\tempvecred,\parametervec,\timevar)\dot{\tempvecred}
    &=
      \stfmat_\text{r}(\tempvecred,\parametervec,\timevar) \tempvecred
      +\cnstvec_\text{r}(\tempvecred,\parametervec,\timevar)+\statenoise,&&t>0,\quad\tempvecred(0)=\tempvecred_0\\
    \measvec_\text{x}
    &=\measmat_\text{r}(\timevar)\tempvecred+\measnoise,&&t\geq 0.
  \end{align}
\end{subequations}
In view of the implementation of the estimator in a process control
unit the Kalman filter is subsequently based on the discrete time
version of system \eqref{eq:pod_red_system}. Taking into account a
suitable discretization scheme, e.g., the explicit Euler method,
Heun's method or an (explicit) Runge-Kutta scheme, formally the sampled data system 
\begin{subequations}
  \label{eq:red:discrete}
  \begin{align}
    \tempvecred_{\timestep}&=\vec{F}_{\timestep-1}(\tempvecred_{\timestep-1},\parametervec_{\timestep-1},\statenoise_{\timestep-1})
                             ,&&k\geq
                              1,\quad\tempvecred_0=\tempvecred(0)\label{eq:red:discrete:ode}\\
    \measvec_{\text{x},\timestep} &= \measmat_{\text{r},\timestep} \tempvecred_{\timestep}+\measnoise_{\timestep}\label{eq:red:discrete:out}
  \end{align}
\end{subequations}
is obtained, where $\tempvecred_{\timestep}=\tempvecred(\timevar_{\timestep})$,
$\parametervec_{\timestep}=\parametervec(\timevar_{\timestep})$,
$\statenoise_{\timestep}=\statenoise(\timevar_{\timestep})$,
$\measnoise_{\timestep}=\measnoise(\timevar_{\timestep})$,
$\timevar_{\timestep}=\sum_{j=1}^{k}\stepsize_j$, $\timevar_0=0$, and $\stepsize_{\timestep}$
is the stepsize or sampling time, respectively, which is assumed to be
distributed unevenly for reasons explained below.
\begin{remark}
  \label{rem:timestep}
  The number of discretization steps for the forming process is fixed
  to an integer $n_t$. Hence each step describes a particular snapshot of
  the forming process determined from the LS-DYNA simulation. In other
  words, instead of performing a multitude of simulations for
  different punch speed $\punchspeed$ and holding time
  $\holdingtime$ we address the resulting effects by adjusting the
  time step $\stepsize_\timestep$. If the forming speed is increased,
  then the step size $\stepsize_\timestep$ is reduced. Accordingly,
  $\stepsize_\timestep$ is increased when the holding time
  $\holdingtime$ is increased. By this, the resulting effect on the
  temperature evolution is reflected accordingly.
\end{remark}

Due to the nonlinear system description an \gls{ekf} is considered,
which is based on the local linearization of \eqref{eq:red:discrete}
with respect to the current state estimate. For this let
$\statetransmat_{\timestep}(\tempvecred_{\timestep})$ and $\jacFw_{\timestep}(\tempvecred_{\timestep})$ denote the 
Jacobian matrices of $\vec{F}_{\timestep}(\tempvecred_{\timestep},\parametervec_{\timestep},\statenoise_{\timestep})$
with respect to $\tempvecred_{\timestep}$ and
$\statenoise_{\timestep}$, respectively, i.e. 
\begin{align}
  \label{eq:ekf_transition}
  \statetransmat_{\timestep}(\tempvecred_{\timestep})=\frac{\partial
  \vec{F}_{\timestep}(\tempvecred_{\timestep},\parametervec_{\timestep},\statenoise_{\timestep})}{\partial
  \tempvecred_{\timestep}}\bigg\vert_{\statenoise_{\timestep}=\vec{0}},\quad
  \jacFw_{\timestep}(\tempvecred_{\timestep})=\frac{\partial
  \vec{F}_{\timestep}(\tempvecred_{\timestep},\parametervec_{\timestep},\statenoise_{\timestep})}{\partial
  \statenoise_{\timestep}}\bigg\vert_{\statenoise_{\timestep}=\vec{0}}.
\end{align}
The linear Kalman filter design can be divided into two steps, namely
the prediction and the update step \citep{Anderson2012}. This similarly
extends to the \gls{ekf} design by making use of the above mentioned
local linearization \citep{gelb:74}. 
In the prediction step the state estimate
$\tempvecredest_{\timestep}^{-}$ is updated based on the previous state
estimate $\tempvecredest_{\timestep-1}$ and parameter vector
$\parametervec_{\timestep-1}$ using \eqref{eq:red:discrete} and the
covariance matrix $\covarmat_{\timestep}^{-}$ is computed
\begin{subequations}
  \label{eq:ekf:2steps}
  \begin{align}
    \tempvecredest_{\timestep}^{-}
    &=
      \vec{F}_{\timestep-1}\big(\tempvecredest_{\timestep-1},\parametervec_{\timestep-1},\vec{0}\big)\\
    \covarmat_{\timestep}^{-}
    & =
      \statetransmat_{\timestep-1}(\tempvecredest_{\timestep-1})
      \covarmat_{\timestep-1}
      \statetransmat_{\timestep-1}^T(\tempvecredest_{\timestep-1})+
      \jacFw_{\timestep-1}(\tempvecredest_{\timestep-1})\statenoisemat\jacFw^T_{\timestep-1}(\tempvecredest_{\timestep-1}).      
  \end{align}
  In the update step the state estimation is updated based on the
  current measurements $\measvec_{\text{x},\timestep}$. At first the
  Kalman gain $\kalmanmat_\timestep$ is defined
  \begin{align}
    \kalmanmat_\timestep
    &=
      \covarmat_{\timestep}^{-}
      \measmat_{\text{r},\timestep}
      \left(\measmat_{\text{r},\timestep} \covarmat_{\timestep}^{-} \measmat_{\text{r},\timestep}^T+\measnoisemat\right)^{-1},
  \end{align}
  which is used to update state estimate and covariance matrix according to 
  \begin{align}
    \tempvecredest_{\timestep}
    &=
      \tempvecredest_{\timestep}^{-}+\kalmanmat_{\timestep}\left(\measvec_{\text{x},\timestep}-
      \measmat_{\text{r},\timestep}\tempvecredest_{\timestep}^{-}\right)\\
    \covarmat_{\timestep}&=\left(\iden-\kalmanmat_\timestep \measmat_{\text{r},\timestep}\right)\covarmat_{\timestep}^{-}.
  \end{align}
\end{subequations}
The state vector $\tempvecredest_{\timestep}$ of the EKF is considered
as the estimate of the state $\tempvecred_{\timestep}$ of the
\gls{rom} defined in \eqref{eq:red:discrete}. By making use of
\eqref{eq:pod_red_state} the estimate
\begin{align}
  \label{eq:ekf:qest}
  \tempvecest_{\timestep}=\projectionleft\tempvecredest_{\timestep}
\end{align}
of the nodal temperature vector of the discrete time version of full order system
\eqref{eq:heat_equation_matrix} is obtained, which is utilized
subsequently for the reconstruction of the temperature distribution in
the sheet metal blank.
\subsection{Property estimation}\label{subsec:property_estimation}
Microstructural properties such as hardness of the product essentially depend on the
deformation and temperature history. Based on the previously
introduced state estimation approach and assuming that the deformation
history is determined by the \lsdyna simulation at a sufficient level
of accuracy, then properties may be formally represented in the form
\begin{align}
  \label{eq:properties}
  \propvec = \propfun (\temperature,\parametervec,\timevar),
\end{align}
where $\propvec(\nodeloc,\timevar)$ summarizes the spatial-temporal
property distribution. The function
$\propfun(\temperature(\nodeloc,\timevar),\parametervec(\nodeloc,\timevar,\controlvec),\timevar)$
may include also the temperature history, which can be represented as
a convolution integral over the temperature distribution with an
appropriate integral kernel. 
In view of the finite element approximation implying \eqref{eq:heat_equation_matrix}, the determination of the
\gls{rom} \eqref{eq:pod_galerkin}, and the state estimation
\eqref{eq:ekf:2steps} the property equation \eqref{eq:properties} can
be re-written in the discrete time form
\begin{align}
  \label{eq:properties:fem:est}
  \propvecest[\timestep] = \propfunest_{\timestep}(\tempvecest_{\timestep},\parametervec_{\timestep})
\end{align}
with $\propvecest[\timestep]$ referring the property distribution
vector at the nodes of the mesh and $\tempvecest_{\timestep}$ the
temperature estimation \eqref{eq:ekf:qest}. 

\subsection{Model simplification using supporting trajectories}
\label{subsec:mode_linearization_around_supporting_trajectory}

The parameter vector $\parametervec(\nodeloc,\timevar,\controlvec)$
in~\eqref{eq:heat_equation_matrix} is obtained from the \lsdyna
simulation, which is performed offline. Since $\parametervec(\nodeloc,\timevar,\controlvec)$ depends
on the inputs \eqref{eq:process:inputs} by means of $\controlvec(\timevar)$ full
scenario coverage would require to numerically solve all possible
combinations offline. To avoid this rather tedious procedure
subsequently a so-called supporting trajectory labelled
$\bar{\parametervec}\cts$ is defined, which is associated with the
respective temperature trajectory $\temptraj\cts$ to perform a 
linearization of the nonlinear system dynamics
\eqref{eq:heat_equation_matrix} or \eqref{eq:pod_galerkin},
see, e.g., \citet{Rewienski2003}. Hence under the assumption that 
$\|\tempvec\cts-\temptraj\cts\|_2$ is sufficiently small the model 
\eqref{eq:heat_equation_matrix} is subsequently approximated by  
\begin{subequations}
   \label{eq:system_traj}
  \begin{align}
    \masmatb\cts\dot{\tempvec}
    &=
      \stfmatb\cts\tempvec+\cnstvecb\cts,
    && t>0,\quad\tempvec(0)=\tempvec_0\in\realnumber^{\dimstate}\\
    \measvec_\temperature
    &=
      \measmat(\timevar)\tempvec,&&t\geq 0
  \end{align}
\end{subequations}
with
$\masmatb\cts=\masmat(\temptraj(\timevar),\parametertraj(\timevar),\timevar)$,
$\stfmatb\cts=\stfmat(\temptraj(\timevar),\parametertraj(\timevar),\timevar)$,
and
$\cnstvecb\cts=\cnstvec(\temptraj(\timevar),\parametertraj(\timevar),\timevar)$. Note
that \eqref{eq:system_traj} does not correspond to a standard
linearization using Taylor series expansion, which would imply a
linearized model describing the deviation to the supporting
trajectory, but rather refers to a model simplification. One
particular advantage of this problem formulation is that the arising
matrices, though dependent on the known variables
$(\temptraj\cts,\bar{\parametervec}\cts)$, can be pre-computed so that
\eqref{eq:system_traj} can be considered as a linear time-varying
system. 
To determine the corresponding \gls{rom} again \gls{pod} can be
applied as introduced in
Section~\ref{subsec:model_order_reduction_using_pod}. This yields the
\gls{rom}
\begin{subequations}
  \label{eq:ltv_system}
  \begin{align}  
    \masmatb_\text{r}\cts\dot{\tempvecred}
    &=
      \stfmatb_\text{r}\cts\tempvecred+\cnstvecb_\text{r}\cts,&&t>0,\quad\tempvecred(0)=\tempvecred_0\in\realnumber^{\dimstatered}\\
    \measvec_\temperature
    &=
      \measmat_\text{r}\cts\tempvecred,&&t\geq 0
  \end{align}
\end{subequations}
with the matrices $\masmatb_\text{r}\cts =\projectionleft^T\masmatb\cts \projectionleft$, $
\stfmatb_\text{r}\cts=\projectionleft^T \stfmatb\cts\projectionleft$,
$\cnstvecb_\text{r}\cts = \projectionleft^T\cnstvecb\cts$, and
$\measmat_\text{r}\cts = \measmat\cts\projectionleft$ for
$\tempvec\cts=\projectionleft\tempvecred\cts$. The state and property
estimation considered in Sections \ref{subsec:ekf} and
\ref{subsec:property_estimation} can be identically evaluated using
\eqref{eq:ltv_system}. To address the resulting errors from the model
approximation as well as neglected physical effects such as phase
transformation in the material a disturbance model is added and
integrated into to the estimator design. 

\subsection{Disturbance estimation}

Unmodeled dynamical effects and approximation/simplification errors
are subsequently
subsumed in terms of an unknown disturbance vector $\dist\cts$ so that
\eqref{eq:heat_equation_matrix} is amended according to
\begin{align}
  \label{eq:heat_equation_matrix:dist}
  \masmat(\tempvec,\parametervec,\timevar)\tempvecdot=\stfmat(\tempvec,\parametervec,\timevar)
  \tempvec+\cnstvec(\tempvec,\parametervec,\timevar)+\distmat(\tempvec,\parametervec,\timevar)\dist. 
\end{align}
The disturbance $\dist\cts\in\realnumber^{\dimdist}$ can be considered as vector of power
densities that act on the system and the disturbance matrix
$\distmat(\tempvec,\parametervec,\timevar)\in\realnumber^{\dimstate\times\dimdist}$
quantifies, where and how this 
power is induced. In particular
$\distmat(\tempvec,\parametervec,\timevar)\dist\cts$ can be used to  
represent the heat induced by phase transformation, whose mathematical
description might be difficult to formulate and to parametrize, using
an external model determining $\dist\cts$. The main issue in
this setup is the proper determination of the disturbance matrix,
which will be illustrated in Section \ref{sec:simulation_results} for
the example of a hole-flanging process by relying on physical
intuition and numerical studies. 

As before a \gls{rom} can be determined for
\eqref{eq:heat_equation_matrix:dist} using \gls{pod}, where the
snapshot matrix \eqref{eq:snapshotmatrix} may be extended by varying
both $\controlvec\cts$ and the disturbance vector $\dist\cts$ to
excite the relevant system dynamics. Proceeding as before and taking
into account the previously introduced approximation using a
supporting trajectory the corresponding \gls{rom} can be represented
as
\begin{align}
  \label{eq:ltv_reduced}
  \masmatb_\text{r}(\timevar)\dot{\tempvecred}(\timevar)&=\stfmatb_\text{r}(\timevar)\tempvecred(\timevar)+\cnstvecb_\text{r}(\timevar)+\distmatb_\text{r}(\timevar)\dist,&&t>0,\quad\tempvecred(0)=\tempvecred_0\in\realnumber^{\dimstatered}. 
\end{align}
The unknown disturbances $\dist\cts$ are estimated online using the
concept of \citet{Meditch1973}. For this, a quasi-static disturbance
model is introduced in the form $\vec{0}=\mathbf{A}_\text{d}\dist\cts$
to augment \eqref{eq:ltv_reduced}. The term quasi-static refers to the
fact that the rate of change of the disturbance can be neglected
compared to the rate of change of the system state. Of course other
choice can be made depending on the process characteristics.
The resulting extended system reads
\begin{subequations}
  \label{eq:ekf_augmented}
  \begin{align}
    \begin{bmatrix}
      \masmatb_\text{r}(\timevar) & {0} \\
      {0} &\iden
    \end{bmatrix}
            \begin{bmatrix}
              \dot{\tempvecred} \\ 
              \dot{\dist}
            \end{bmatrix}
                                  &=
		\begin{bmatrix}
		\stfmatb_\text{r}(\timevar) & \distmatb_\text{r}(\timevar)\\
		{0} & \mathbf{A}_\text{d}\cts
		\end{bmatrix}
	\begin{bmatrix}
	\tempvecred\\
	\dist
      \end{bmatrix}
    +
    \begin{bmatrix}
      \cnstvecb_\text{r}(\timevar)\\\mathbf{0}
    \end{bmatrix}
    +\statenoise(\timevar),&&t>0,\quad
                              \begin{bmatrix}
                                \tempvecred\\
                                \dist
                              \end{bmatrix}(0) =   \begin{bmatrix}
	\tempvecred_0\\
	\dist_0
      \end{bmatrix}                       
    \\
    \measvec_\text{x}&=
                       \begin{bmatrix}
                         \measmat_\text{r}(\timevar) & \mathbf{0}
		\end{bmatrix}
	\begin{bmatrix}
	\tempvecred\\
	\dist
	\end{bmatrix}+\measnoise(\timevar).
  \end{align}
\end{subequations}
The \gls{ekf} design from Section~\ref{subsec:ekf} can then be applied
to \eqref{eq:ekf_augmented} with the process noise covariance matrix
$\statenoisemat$ amended by the one $\statenoisemat_\text{d}$ of the
disturbance model. This results in a combined state, property and 
disturbance estimation based on the augmented \gls{rom}. Note that
due to the injection of the measurement and the correction of the
augmented state estimate $\hat{\vec{x}}_e\cts =
[\tempvecredest^{T}\cts,\hat{\dist}\cts]^T$ by the Kalman gain 
the disturbance estimate $\hat{\dist}\cts$ is adjusted despite the
assumption of quasi-stationarity of the disturbance model.

\section{Simulation results}
\label{sec:simulation_results}

While the derived methods are generally applicable to multi-stage
forming processes they are subsequently evaluated in simulation
studies for a hole-flanging process. This process is first described
and analysed before the model and approximation errors are
quantified. State estimation is validated for two scenarios. First, it
is evaluated based on a simulation using the finite element code
Firedrake with known disturbances. Secondly, the solution of the
\lsdyna simulation involving phase transformation is used to feed the
state and disturbance estimation based on the derived \gls{rom}.

\subsection{Hole-flanging process}
\label{subsec:collar_forming_as_example_process}

\begin{figure}[!ht]
  \centering
  \includegraphics{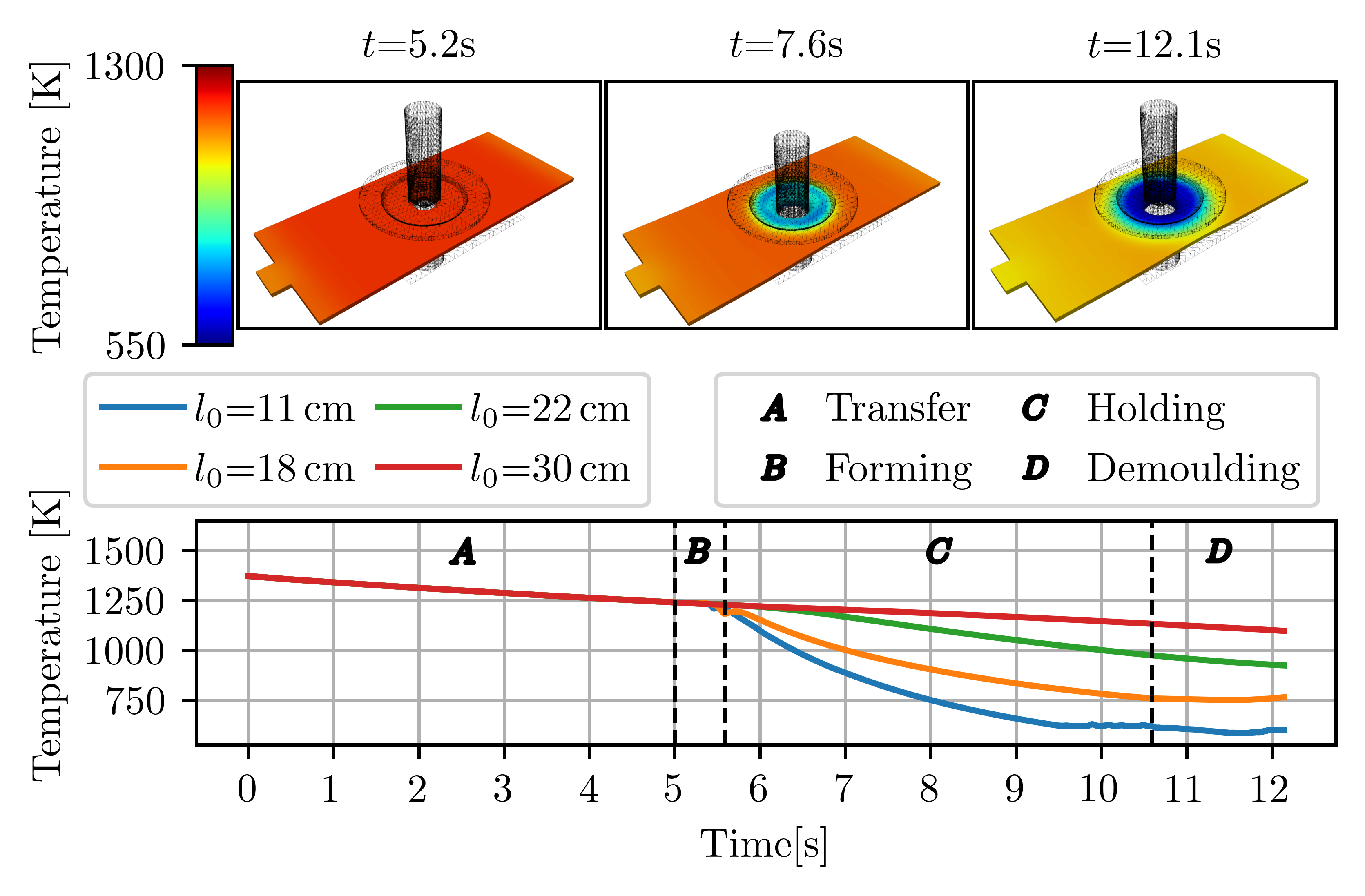}
  \caption{Temperature snapshots during the forming process obtained
    from a full order coupled thermo-mechanical simulation using
    \lsdyna (top). Temperature evolution at four different
    locations with different distances $l_0$ to the centre of the
    sheet metal blank (bottom). The effect of the phase
    transformation from austenite to martensite can be seen at
    approximately \SI{10}{\second}.}
  \label{fig:lsdyna_overview}
\end{figure}

The validation takes place for a hole-flanging process in
simulation. An illustration is shown in
Figure~\ref{fig:lsdyna_overview}. The forming process is separated in
four phases: transfer (A), forming (B), holding (C), and demoulding
(D). Three snapshots are taken before the forming process, during
holding in the lower dead point and after the forming process at
$t=\SI{5.2}{\second}$, $t=\SI{7.6}{\second}$ and
$t=\SI{12.1}{\second}$, respectively. The bottom graph shows the 
temperature evolution at different nodes having a certain distance
$l_0$ to the centre point of the sheet metal blank obtained from a
full order coupled thermo-mechanical simulation using \lsdyna. The
regions having tool contact ($l_0=\SI{11}{\centi\metre}$ and
$l_0=\SI{18}{\centi\metre}$) cool down most, while regions which are outside the tool contact region
($l_0=\SI{22}{\centi\metre}$ and $l_0=\SI{30}{\centi\metre}$) are less
affected by the forming process.
At the end of the holding phase, it can be seen that the point with a
distance of \SI{11}{\centi\metre} stops cooling down and holds the
temperature for some time. This effect is caused by the phase
transformation.

The full order thermal model \eqref{eq:heat_equation_matrix} is of
dimension $\dimstate=\num{17181}$. The input vector $\controlvec(\timevar)$ in this
process contains the punch speed \punchspeed and the austenite
temperature \tempaustavg. The simulation is composed of 510
discretization steps $n_t$ and the step sizes $\stepsize_\timestep$
are chosen according to Remark \ref{rem:timestep}. It
is assumed that pointwise temperature measurements
$\measvec_\temperature(\timevar)$ fixed to certain positions in the world
frame are available leading to the time-variant output matrix
$\measmat(\timevar)$ in~\eqref{eq:heat_output}. Shadowing of the
sensors is not considered and measurement noise $\measnoise(\timevar)$
with a standard deviation of \SI{10}{\kelvin} is added to the signals
to evaluate the performance of the state estimation. The supporting
trajectory $(\bar{\parametervec}\cts,\bar{\tempvec}\cts)$ is obtained from the \lsdyna
simulation with $\tempaust=\SI{1273}{\kelvin}$ and
$\punchspeed=\SI{80}{\milli\meter\per\second}$. 

\begin{figure}[!ht]
  \centering
  \includegraphics{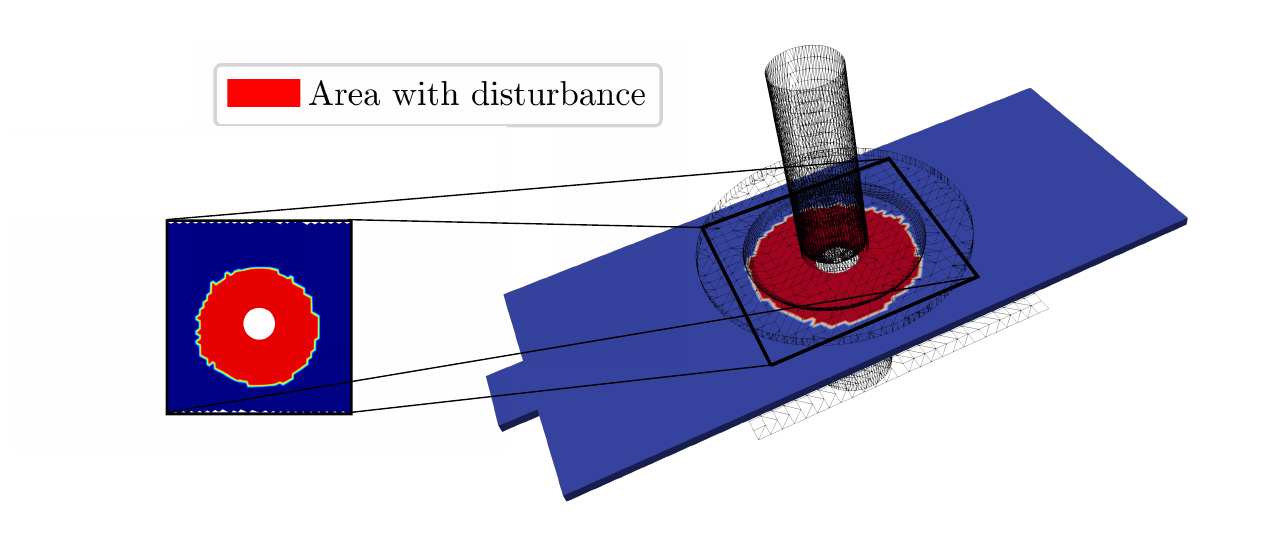}
  \caption{Red colored areas illustrate the region of phase
    transformation at tool contact and hence the domain
    addressed primarily by the disturbance model.} 
  \label{fig:dist_model}
\end{figure}

The disturbance model is physically motivated by the phase
transformation from austenite to martensite. In the given process, it
assumed that the transformation takes place simultaneously at all
nodes that have tool contact, which is illustrated by the red area in
Figure~\ref{fig:dist_model}. It is assumed that there is no phase
transformation in the blue area during the simulation time. 

\subsection{Model order reduction}
\label{subsec:model_order_reduction}

An important step in the POD reduction method is the choice of the
snapshot matrix. For the given example it is a concatenation of
several full order simulations with arbitrary excitation of the
disturbance $\dist\cts$ and varying inputs $\controlvec\cts$ to the
system. The austenitizing temperature $\tempaust$ is varied between
\SI{1073}{\kelvin} and \SI{1373}{\kelvin} while the punch speed
\punchspeed is varied between \SI{80}{\milli\metre\per\second} and
\SI{100}{\milli\metre\per\second}.  

\begin{figure}[!ht]
  \centering
  \includegraphics{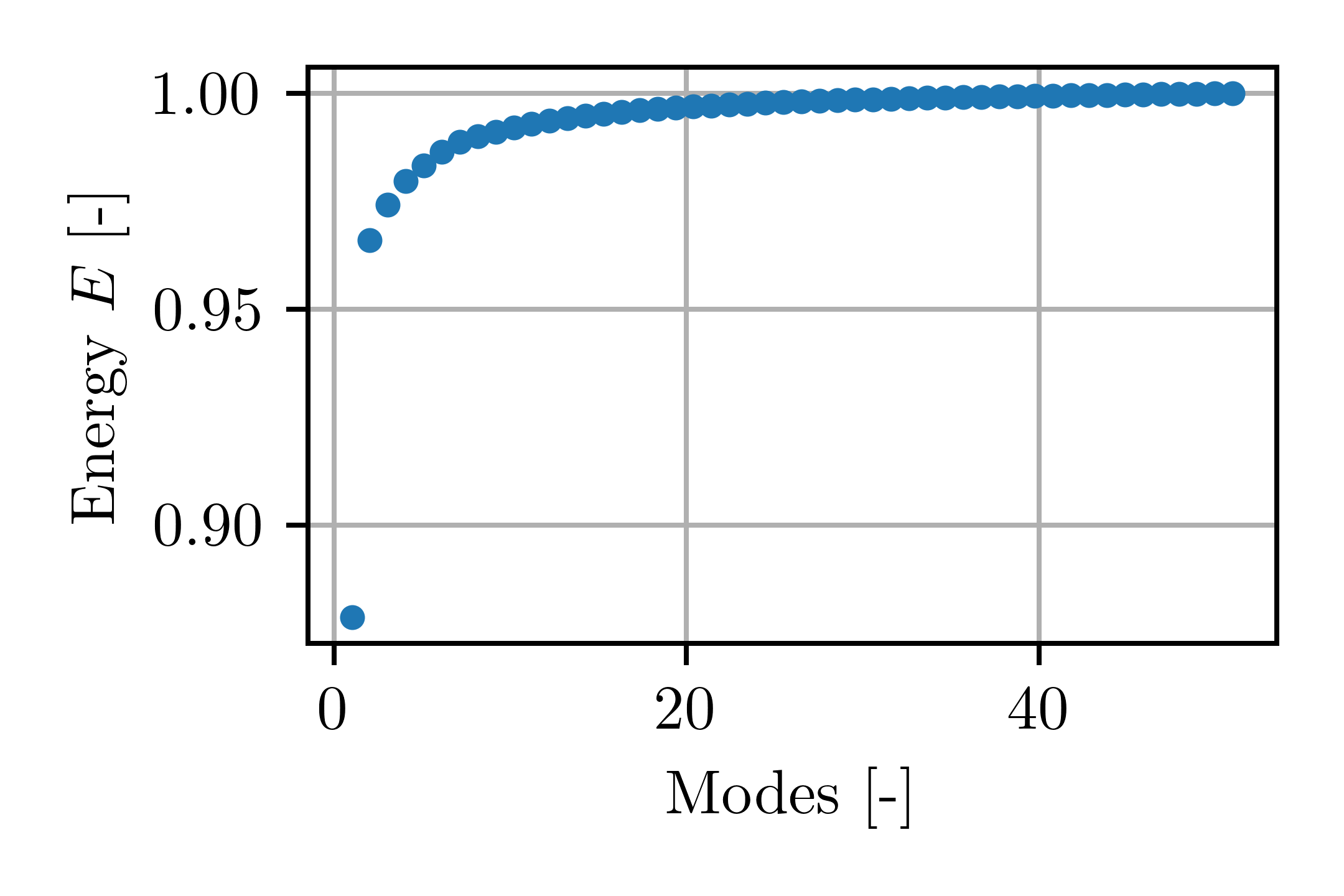}
  \caption{Approximated cumulative energy \eqref{eq:pod_energy}
    covered by the \gls{pod} modes.} 
  \label{fig:pod_svd}
\end{figure}

In Figure~\ref{fig:pod_svd}, the energy \podenergy of the snapshots
that is covered by the reduced system with \dimstatered modes as
described in~\eqref{eq:pod_energy} is shown. In the given case
$\dimstatered=30$ modes cover over 99 \% of the energy of the snapshot
matrix.
\begin{figure}[!ht]
  \centering
  \includegraphics[width=\textwidth]{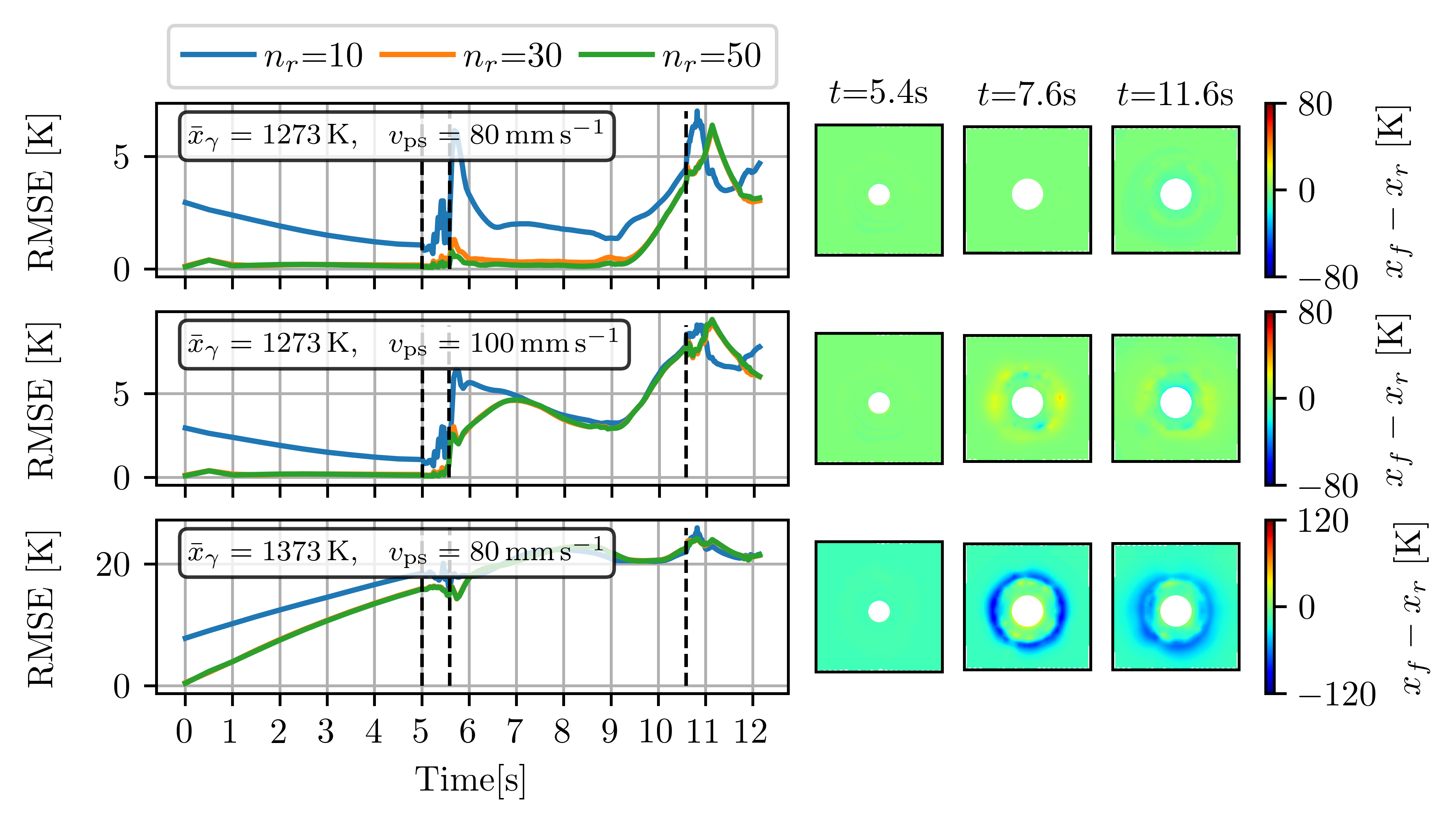}
  \caption{RMSE \eqref{eq:rmse} for input variations in terms of
    austenitizing temperature \tempaustavg and punch speed \punchspeed
    (left column) and corresponding surface plots to illustrate
    spatial distribution of the modeling error (right column) for
    different dimensions $\dimstatered\in\{10,30,50\}$ of the \gls{rom}.}
  \label{fig:pod_overview}
\end{figure}
This assumption is also confirmed in Figure~\ref{fig:pod_overview},
where the \gls{rmse} between the full model and the reduced order
model is shown for three simulations. The \gls{rmse} is thereby
computed at each time step $\timestep$ as 
\begin{align}
  \label{eq:rmse}
  e_\timestep=\frac{\int_{\volume_\timestep}||\projectionleft{\tempvecred}_\timestep-\tempvec_\timestep||_2
  d\volume}{\int_{\volume_\timestep} d\volume}, 
\end{align}
where $\projectionleft{\tempvecred}_\timestep$ is the reconstructed
sheet metal temperature based the reduced order states $\tempvecred_\timestep$
using \eqref{eq:pod_red_state}.
In the simulation with $\tempaustavg=\SI{1273}{\kelvin}$ and
$\punchspeed=\SI{80}{\milli\metre\per\second}$ the error converges to
\SI{0}{\kelvin} as $\dimstatered$ is increased. This is due to the
fact that this setting corresponds to the supporting trajectory $(\bar{\parametervec}\cts,\bar{\tempvec}\cts)$ that is used for the
model simplification in~\eqref{eq:ltv_system}. From \SI{9}{\second} to
\SI{11}{\second} a disturbance with a power density of
\SI{1000}{\watt\per\cubic\metre} is applied to the system. As this
particular disturbance signal is not present in the snapshot matrix
$\snapshotmatrix$, it leads to a certain error. In the surface plots
in the right column of Figure \ref{fig:pod_overview}, which show the
same subset as in Figure~\ref{fig:dist_model}, it can be seen that the
error of the disturbance arises mainly around the 
area with tool contact. In the second and third row, the
simplification error takes effect. The \gls{rmse} with respect to a variation in
\punchspeed is below \SI{1}{\kelvin} in the transfer phase because the
parameter trajectory is the same as in the nominal case. The
\gls{rmse} rises to around \SI{5}{\kelvin} in the forming phase, where
the parameter and temperature trajectories deviate from the supporting
trajectories. The surface plots illustrate that this error is dominant
in the area of tool contact. The \gls{rmse} for a variation in
\tempaustavg grows up to \SI{20}{\kelvin} without the influence of
disturbances. The surface plots illustrate that there is also an error
outside the tool contact region which is mainly caused by the
temperature dependent material properties in~\eqref{eq:heat_eq_pde}.

The computational time for the simulation of the hole-flanging
process is reduced from \SI{33000}{\second} in \lsdyna to \SI{405}{\second} for the full order model
\eqref{eq:heat_equation_matrix} to \SI{0.3}{\second} for the \gls{rom}
\eqref{eq:ltv_reduced} with $\dimstatered=30$ states using the
introduced model simplification in terms of the supporting
trajectory. This implies a reduction in computation time by a factor
of 1000  from the full order model and confirms the real-time capabilities of the derived
\gls{rom}. The simulation is conducted on a laptop with an Intel
Core i5-8265U with \SI{2.8}{\giga\hertz}.

\subsection{State and disturbance estimation with full order thermal
  plant model}\label{subsec:sim:full}

The state and disturbance estimation is first evaluated in combination
with the full order thermal model \eqref{eq:heat_equation_matrix}. A
disturbance signal $\dist\cts$ with a power density of 
\SI{1000}{\watt\per\cubic\metre} between \SI{9}{\second} and
\SI{11}{\second} is imposed to this model, which is unknown to the
\gls{ekf} and needs to be reconstructed utilizing the local
temperature measurements. The austenitizing temperature \tempaustavg and punch
speed \punchspeed are assigned as \SI{1373}{\kelvin} and
\SI{80}{\milli\metre\per\second}, respectively.

\begin{figure}[!ht]
  \centering
  \includegraphics[width=\textwidth]{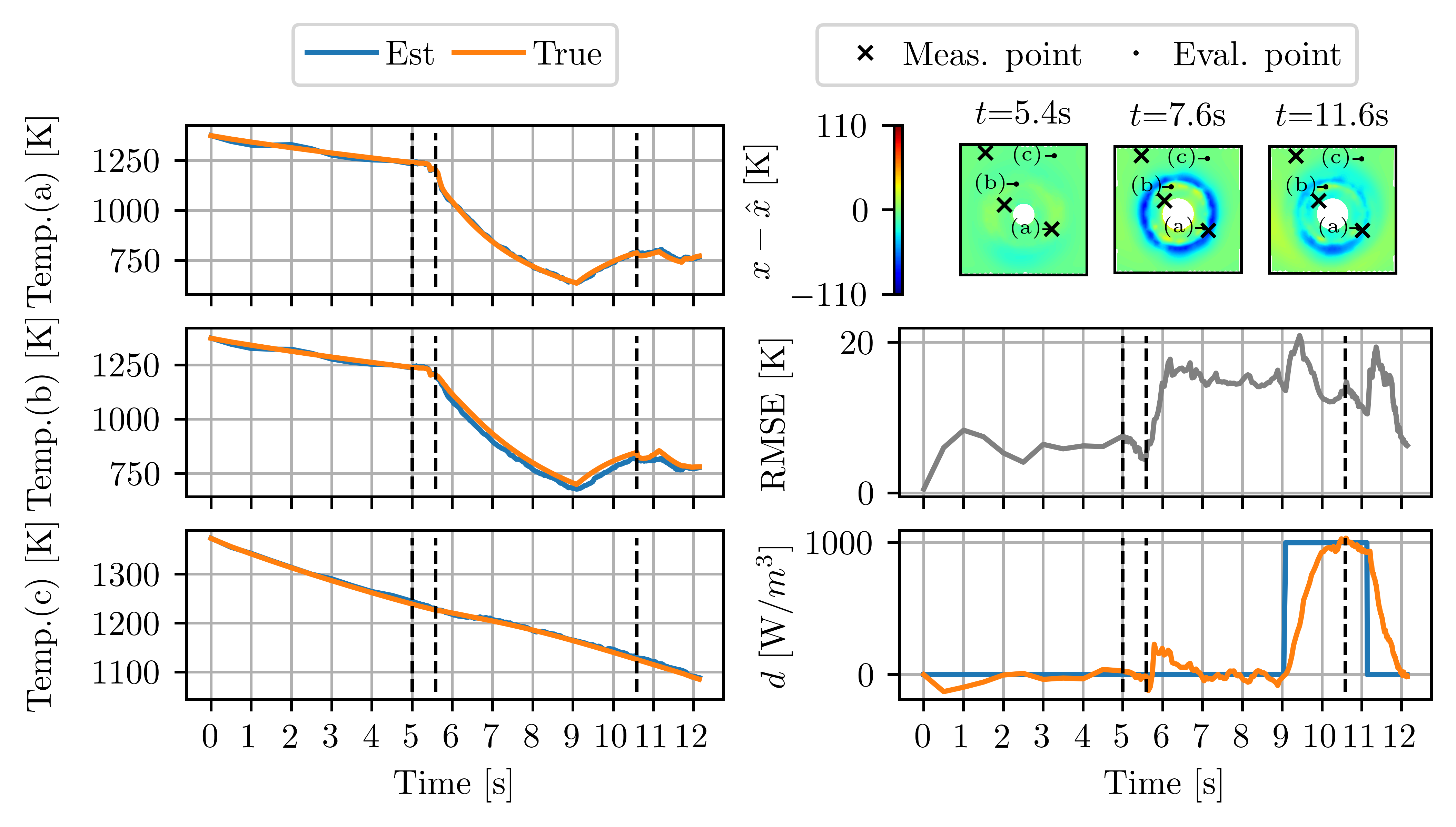}
  \caption{State and disturbance estimation using the \gls{ekf} based
    on the augmented \gls{rom} \eqref{eq:ekf_augmented} with the full
    order thermal model \eqref{eq:heat_equation_matrix} of Firedrake serving as plant
    model for austenitizing temperature $\tempaustavg=\SI{1373}{\kelvin}$ and 
    punch speed $\punchspeed=\SI{80}{\milli\metre\per\second}$. Left
    column: comparison of estimated and simulated temperature at 
    evaluation points; right column: location of measurement and
    evaluation points (top), RMSE (middle) and comparison between
    applied and estimated disturbance (bottom).}
  \label{fig:ekf_full_overview}
\end{figure}

The \gls{ekf} in Section~\ref{subsec:ekf} needs to be initialized with
the covariance matrices of process and sensor noise. Subsequently,
diagonal matrices are assumed with
$\statenoisemat=10\iden_{\dimstatered,\dimstatered}$ and
$\measnoisemat=\num{0.1}\iden_{\dimmeas,\dimmeas}$, where
$\iden_{i,j}$ is the $i\times j$ identity matrix. The initial state
of the \gls{ekf} is assigned as $\tempvecredest_0=\projectionleft^{T}\tempvec_0$ and hence
corresponds to the projection of the systems initial state. The
initial value of the covariance matrix is chosen as
$\covarmat_{0}=10\iden_{\dimstatered,\dimstatered}$. 
The \gls{ekf} is driven by the process input $\controlvec(\timevar)$ and
$\dimmeas=3$ temperature measurements summarized 
in the vector $\measvec_\temperature(\timevar)$. Their location is indicated in
Figure~\ref{fig:ekf_full_overview}, top right column by the crosses in the surface plots. The measurements are subsequently
computed from the full order model
\eqref{eq:heat_equation_matrix}. Note that sensor position is assumed
spatially fixed so that the temperature of different nodes passing the
measurement location due to the deformation is extracted by the matrix
$\measmat(\timevar)$. 

The obtained estimation results are summarized in
Figure~\ref{fig:ekf_full_overview}. In the left column the temperature
evolution for selected evaluation points on the sheet metal blank
(corresponding to the numbers in the surface plots) are given. The
first point (a) is located on a measurement point inside 
the disturbance region. In this case, the estimated temperature
follows the true temperature without a large deviation because the
disturbance estimate is chosen such that the error is minimized. In
the bottom right plot, the estimated disturbance signal is compared
with the applied disturbance signal. The estimate follows the true
trajectory with a small delay because a constant disturbance in the model
in~\eqref{eq:ekf_augmented} is assumed. The second point (b) is not
placed on a measurement point so that the temperature evolution is
reconstructed by the \gls{ekf}. Due to modeling errors the temperature is not 
reconstructed as good as in (a). The third node (c) is also not placed
on a measurement point. However, the reconstruction is better than
(b), because it is not placed within the tool contact area so that
the model is more accurate. Overall the state and disturbance
estimation perform well with a \gls{rmse} below \SI{20}{\kelvin}, in
particular in view of the comparatively low number of measurement
values. 

\subsection{State and disturbance estimation with thermo-mechanical
  \lsdyna plant model}

The state estimation is furthermore tested on the \lsdyna simulation,
where the thermal effects of the phase transformation are included by
the internally used material model, so that the disturbance estimator
is used to capture the respective effects. The austenitizing temperature
$\tempaust$ and the punch speed \punchspeed are assigned as
\SI{1373}{\kelvin} and \SI{80}{\milli\meter\per\second},
respectively. The sensor configuration and the initialization of the
\gls{ekf} correspond to those of Section \ref{subsec:sim:full}.

\begin{figure}[!ht]
  \centering
  \includegraphics[width=\textwidth]{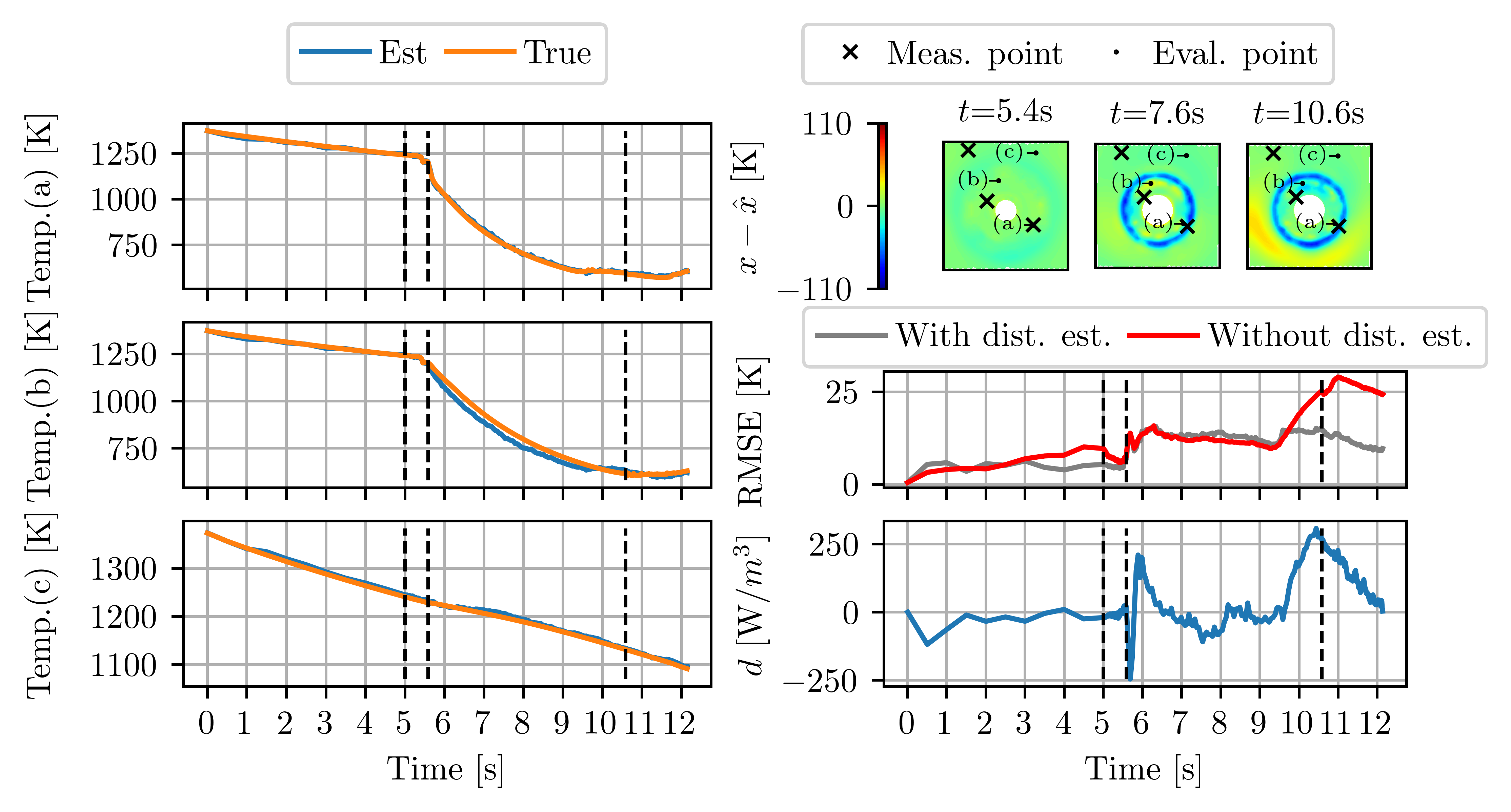}
  \caption{State and disturbance estimation using the \gls{ekf} based
    on the augmented \gls{rom} \eqref{eq:ekf_augmented} with the full
    coupled thermo-mechanical model of \lsdyna serving as plant
    model for austenitizing temperature $\tempaustavg=\SI{1373}{\kelvin}$ and 
    punch speed $\punchspeed=\SI{80}{\milli\metre\per\second}$. Left
    column: comparison of estimated and simulated temperature at 
    evaluation points; right column: location of measurement and
    evaluation points (top), RMSE with and without disturbance
    estimation (middle) and estimated disturbance (bottom).} \label{fig:ekf_lsdyna_overview}
\end{figure}

In Figure~\ref{fig:ekf_lsdyna_overview} state and disturbance
estimation are evaluated compared to the fully coupled
thermo-mechanical \lsdyna simulation, which is used to generate the
measurement signals. As before the estimation results are more
accurate for regions outside the tool contact area, subsequently
represented by evaluation point (c), and at the measurement points,
here evaluation point (a). In overall compared to the results of
Section \ref{subsec:sim:full} the \gls{rmse} is slightly increased
reaching a value of approximately \SI{25}{\kelvin} without and
\SI{15}{\kelvin} with disturbance estimation (cf. Figure
\ref{fig:ekf_lsdyna_overview}, right column, middle). The maximal
\gls{rmse} rises when the phase transformation occurs, which is in the
interval between \SI{9}{\second} and \SI{11}{\second}. The disturbance
estimation (right column, bottom) herein reaches an approximate power density of
\SI{250}{\watt\per\cubic\metre}. In addition, two additional peaks of
$\pm$\SI{250}{\watt\per\cubic\metre} emerge at the beginning
of the holding phase. These are unrelated with the phase transformation
and occur due to other modeling inaccuracies.

\section{Conclusion}
\label{sec:conclusion}

Model-based state and disturbance estimation is developed using
reduced-order models for a multi-stage hot sheet metal forming
process. For this, based on the assumption of a separation between the
mechanical solution and the thermal solution of the thermo-mechanical
process, a nonlinear parametric model describing the spatial-temporal
evolution of the temperature in the sheet metal blank during the
forming process is derived taking into account a finite element
approximation of the heat equation on the time-varying deforming
volume. To address the high model order proper orthogonal
decomposition is utilized to determine a reduced-order model capturing
the essential dynamics of the thermal sub-process. Based on this
system description an Extended Kalman filter is designed and augmented
by a disturbance model to address modeling and approximation errors as
well as neglected physical effects such as phase transformation. The
computational complexity is further reduced by considering a
suitable supporting trajectory which enables us to approximately
consider an linear time-varying system dynamics. The performance of
the state and disturbance estimation is evaluated in simulation
studies for a hole-flanging process coupling the estimator with first the full order thermal model
and secondly with a fully coupled thermo-mechanical simulation using
\lsdyna. The obtained results confirm the performance of the
estimation approach and underline its ability to reconstruct the
spatial-temporal temperature distribution in the sheet metal blank
during the forming process in real-time based on only sparse local
temperature measurements.

Current research considers the extension to property estimation as is
already outlined schematically in this paper. Here, both the
deformation history and the estimated temperature distribution
starting at its initial state  will be utilized in conjunction with
property-related characteristic diagrams or material models to
determine spatial-temporal property distributions. Both, temperature
and property estimation will furthermore serve as fundamental
ingredients for the model-based control of multi-stage hot sheet metal forming
processes.

\section*{Acknowledgements}

The authors thank the Deutsche Forschungsgemeinschaft (DFG) for the
financial support in the project 424334660 (Tekkaya/Meurer) within the
Priority Program SPP2183 "Property-controlled forming processes".

\bibliographystyle{elsarticle-harv}
\bibliography{references}
\end{document}